\documentclass[numerical,onecolumn,superscriptaddress,showpacs,floatfix,reprint]{revtex4}

\usepackage{amsmath,amssymb,amsfonts}
\usepackage[colorlinks=true,plainpages=true]{hyperref}

\usepackage[dvipdfm]{graphicx} 
\usepackage{bmpsize}

\begin{document}
\title{
Centrality definition in e+A collisions at the Electron-Ion Collider
}

\medskip

\author{Mariam Hegazy}
\affiliation{Department of Physics, Faculty of Science, Cairo University, Giza 12613, Egypt}

\author{Aliaa Rafaat}
\affiliation{Department of Physics, The American University in Cairo, New Cairo 11835, Egypt}

\author{Niseem~Magdy} 
\email[Author to whom any correspondence should be addressed~]{niseemm@gmail.com}
\affiliation{Department of Physics, University of Tennessee, Knoxville, TN, 37996, USA}

\author{Wenliang~Li}
\affiliation{Department of Physics, Stony Brook University, Stony Brook, NY 11794, USA}
\affiliation{Center for Frontiers in Nuclear Science at SBU, Stony Brook, NY 11794, USA}

\author{Abhay~Deshpande} 
\affiliation{Department of Physics, Stony Brook University, Stony Brook, NY 11794, USA}
\affiliation{Department of Physics, Brookhaven National Laboratory, Upton, New York 11973, USA}
\affiliation{Center for Frontiers in Nuclear Science at SBU, Stony Brook, NY 11794, USA}

\author{A. M. H. Abdelhady}
\affiliation{Department of Physics, Faculty of Science, Cairo University, Giza 12613, Egypt}


\author{A.Y.Ellithi}
\affiliation{Department of Physics, Faculty of Science, Cairo University, Giza 12613, Egypt}




\begin{abstract}
In this work, we investigate the feasibility of defining centrality in electron-ion collisions at the Electron-Ion Collider (EIC) by examining the correlation between the impact parameter and several observables, including total energy, total transverse momentum, and total number of particles. Using the BeAGLE Monte Carlo generator, we simulate e+Au and e+Ru collisions at different energies and analyze the correlation between the impact parameter and these observables across different kinematic regions. Our findings indicate that the correlation is weak in the central rapidity region but becomes stronger in the forward and far-forward rapidity regions. However, the correlation is not sufficiently robust to allow for precise centrality determination. We conclude that defining centrality in electron-ion collisions is more challenging than in ion-ion collisions, necessitating further studies to develop a robust centrality definition for the EIC.
\end{abstract}
\maketitle

\section{Introduction}
In physics, the impact parameter (b) is defined as the perpendicular distance between the asymptotic trajectories of two colliding particles~\cite{Krane:359790}. Simply put, it represents how close the centers of the two particles would have come to each other if they had continued on their initial paths without interacting. In the context of nuclear collisions, the meaning of the impact parameter depends on the target and projectile.

\begin{figure}[!h]
    \centering
    \includegraphics[width=0.99\linewidth,angle=0]{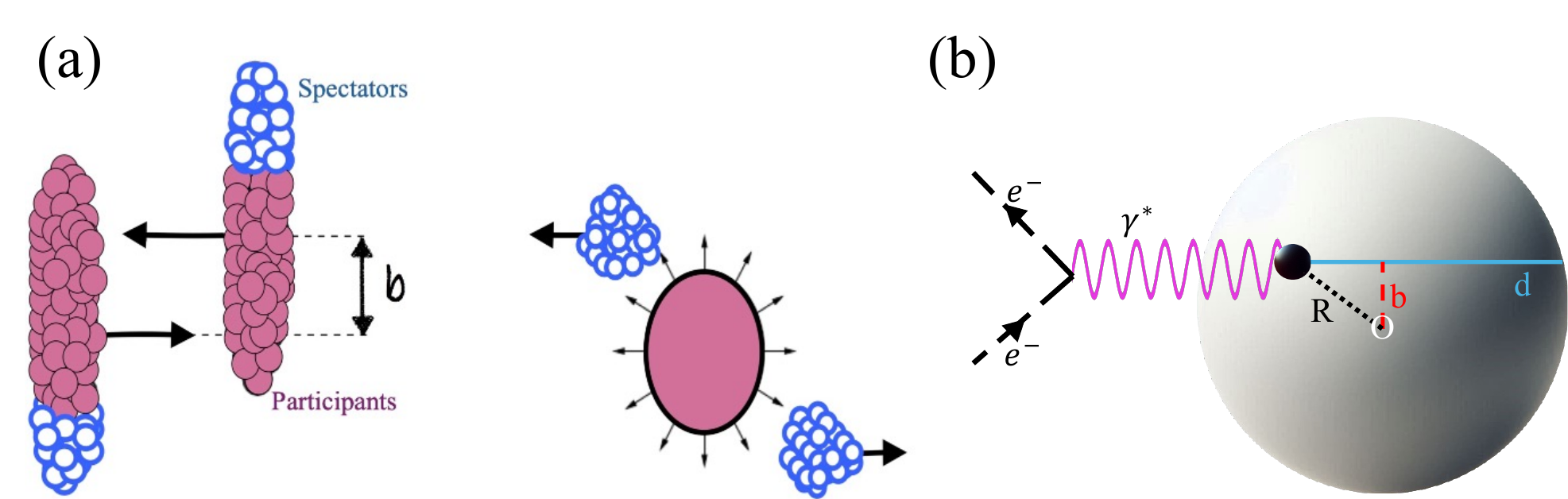}
    \caption{Schematic illustration of a heavy ion collision seen in panels (a). Panel (a) illustrates the definition of the impact parameter and shows the participants and spectators after the collision. Panel (b) shows a schematic illustration of the key quantities to describe the collision geometry, including b, which represents the impact parameter; R, indicating the spatial displacement from the interaction point to the center of the nucleus; and d, defining the projected virtual photon traveling length from the interaction point to the edge of the nuclear medium.}
    \label{fig:0}
\end{figure}

In ion-ion collisions, the impact parameter (i.e., the distance between the centers of the two ions in the plane transverse to the beam axis) is critical for determining the centrality of the collision. The impact parameter defines the overlap region of the nuclei and thus determines the size and shape of the resulting medium. A small impact parameter (see Fig.~\ref{fig:0} panel (a)) signifies a head-on collision with significant overlap of the nuclei, whereas a large impact parameter indicates a peripheral collision with minimal overlap.
Theoretically, the centrality of a nucleus-nucleus collision with impact parameter is usually defined as a percentile in the nucleus-nucleus total cross-section $\sigma_{AA}$~\cite{Miller:2007ri}. 
Additionally, while geometrical quantities like the impact parameter, number of participants ($N_{part}$), number of spectators ($N_{spec}$), and number of collisions ($N_{coll}$) are not directly measurable in experiments, they are correlated with measurable quantities such as charged-particle multiplicity ($N_{Ch}$). Understanding the strength of these correlations is crucial for improving control over centrality selections in ion-ion collisions. In this context, the Glauber model is commonly employed~\cite{Miller:2007ri} to align experimental observables with theoretical predictions and extract the corresponding geometrical properties of the collision.
In ion-ion collision experiments, observables related to the collision geometry, such as average charged-particle multiplicity $N_{ch}$ and the energy carried by particles close to the beam direction that is often deposited in Zero-Degree Calorimeters (ZDC), known as zero-degree energy $E_{ZDC}$, are used for centrality selections~\cite{ALICE:2013hur}.

In e+A collisions (see Fig.~\ref{fig:0} panel (b)), the impact parameter is defined as the transverse distance between the path of the incident photon and the center of the target nuclei~\cite{AbdulKhalek:2021gbh, Chang:2022hkt}. In these collisions, the impact parameter is random, and there is no clear relationship between it and the final observables. Previous studies have indicated that the energy of forward neutrons can be used to define centrality in e+A collisions~\cite{Chang:2022hkt}. However, this raises an important question: "To what extent can the properties of produced particles (i.e., energy and momentum) be used to define centrality in e+A collisions at the EIC?" 
Answering this question would be valuable for studies focused on understanding nuclear structure~\cite{Magdy:2024thf}, as well as for other investigations related to the geometry of e+A collisions~\cite{Li:2023dhb}.

In the 2030s, the Electron-Ion Collider (EIC) is set to revolutionize nuclear deep inelastic scattering by introducing collider kinematics, offering unprecedented opportunities for advancing nuclear physics research~\cite{AbdulKhalek:2021gbh}. Within the framework of e+A collisions at the EIC, the reaction mechanism is divided into three distinct stages~\cite{Chang:2022hkt}: (i) hard scattering processes~\cite{Piller:1995kh}, (ii) IntraNuclear Cascade (INC) processes~\cite{Cugnon:1982qw}, and (iii) the breakup of excited nuclear remnants~\cite{Weisskopf:1937zz}. These stages are thought to occur sequentially on different time scales in the nuclear rest frame. Hard scattering processes occur almost instantaneously at the moment of collision, while INC processes and the subsequent breakup of excited nuclei happen on the order of \(10^{-22}\) and \(10^{-16}\) seconds, respectively~\cite{Serber:1947zz, Monira:2023}. Additionally, the time scales of these processes are expected to vary depending on the properties of the target nuclei and the specifics of the collision kinematics. Therefore, isolating kinematic regions dominated by each process will be crucial for refining nuclear excitation models and gaining deeper insights into the target nucleus~\cite{Mathews:1982zz, Magdy:2024thf}.
Consequently, our work addresses the question raised before in the EIC context by studying the correlations between the impact parameter and several observables, such as the event's total energy $\langle E \rangle$, total transverse momentum $\langle p_{T} \rangle$, and multiplicity $\langle N \rangle$, at different kinematic regions within the framework of the Benchmark e+A Generator for Leptoproduction in high-energy lepton-nucleus collisions (BeAGLE) model.

Here, an important objective is to investigate the ability of the suggested observables to constrain the impact parameter in e+A collisions for different collision energies and system sizes. In this work, we analyze the BeAGLE model events generated for $e+Au$ at 18$\times$110 and 10$\times$40 GeV, and for $e+Ru$ at 18$\times$110 GeV. The paper is organized as follows: Section~\ref{sec:2} summarizes the theoretical model and the observables used in this work, Section~\ref{sec:3} presents the results from the model studies, and Section~\ref{sec:4} provides a summary.

\section{Methodology}\label{sec:2}
\begin{figure}[!h]
 \centering{
 \includegraphics[width=0.99\linewidth,angle=0]{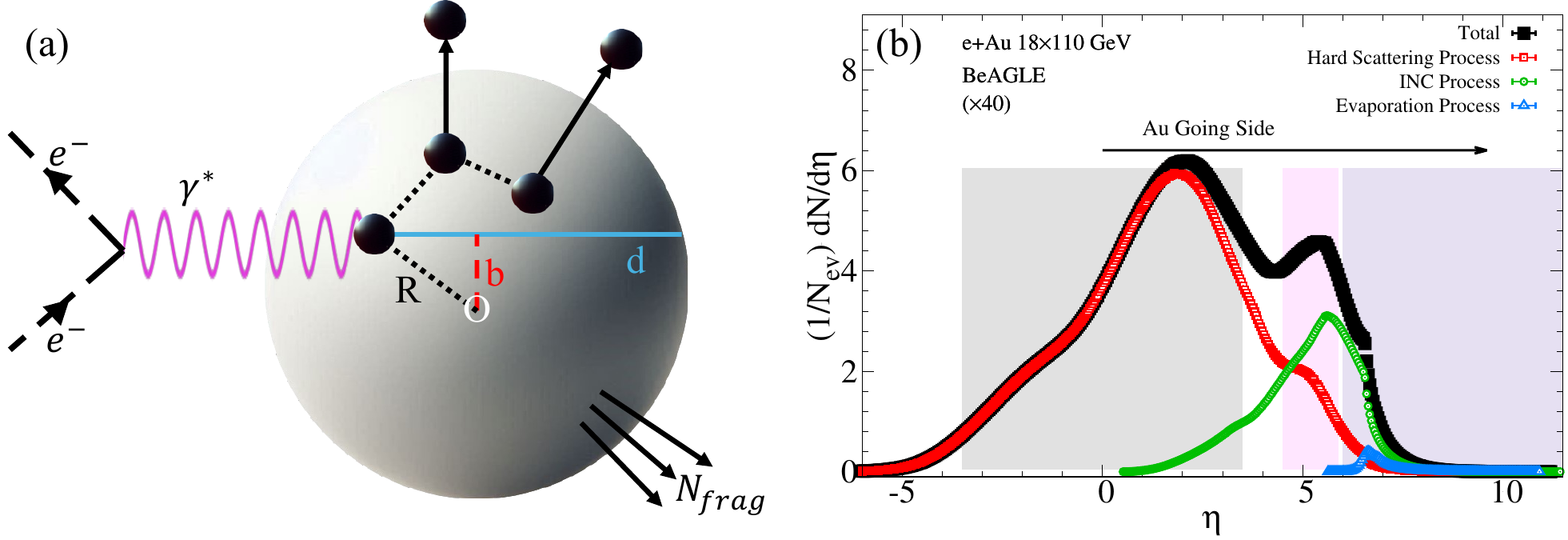}
\vskip -0.36cm
 \caption{
Panel (a) provides a schematic representation of the essential quantities used to describe the e+A collision geometry as given by the BeAGLE model: $b$, representing the impact parameter; $R$, indicating the spatial displacement from the interaction point to the nucleus's center; and $d$, defining the projected length traveled by the virtual photon from the interaction point to the edge of the nuclear medium. Panel (b) illustrates the $dN/d\eta$ distribution of the $e$+$Au$ collisions at 18$\times$110 (GeV) from the BeAGLE model. The shaded areas represent the three acceptances used in this work.
 \label{fig:1}
 }
 }
 \end{figure}

The study utilized version 1.03 of the Monte Carlo (MC) code known as BeAGLE~\cite{Chang:2022hkt}. BeAGLE functions as a hybrid model, integrating several established codes to provide a comprehensive description of high-energy lepton-nucleus (e+A) scattering phenomena (see Fig.~\ref{fig:1} panel (a)). The model combines the following components:
\begin{itemize}
\item{DPMJet} This code models hadron creation and interactions with the nucleus through an intra-nuclear cascade.
\item{PYTHIA6} This model handles partonic interactions and the subsequent fragmentation process.
\item{PyQM} It offers the geometric density distribution of nucleons within a nucleus and incorporates the Salgado-Wiedemann quenching weights to account for partonic energy loss~\cite{SW:2003}.
\item{FLUKA} This code describes the decay of excited nuclear remnants, including nucleon and light ion evaporation, nuclear fission, Fermi breakup of decay fragments, and photon emission de-excitation.
\item{LHAPDF5} Used in conjunction with FLUKA, this model defines high-energy lepto-nuclear scattering.
\end{itemize}
Additionally, BeAGLE provides functionalities for steering, multi-nucleon scattering (shadowing), and an enhanced description of Fermi momentum distributions of nucleons within nuclei.

BeAGLE provides a variety of options for managing simulation phenomena.  This includes the ability to describe nuclear shadowing via various approaches, account for hadron formation time in the DPMJet intranuclear cascade, and tailor the Fermi motion of nucleons within the nucleus through different mechanisms. PyQM functionalities extend to specifying the transport coefficient $\hat{q}$ to modulate the interaction degree between energetic partons and the nuclear environment and fine-tune details of the partonic energy loss process. The main program, DPMJet, interfaces with PYTHIA6 to handle elementary interactions and fragmentation. PyQM, in turn, manages this process directly post-elementary interactions in PYTHIA6, while DPMJet is responsible for nuclear geometry and nuclear evaporation post-fragmentation facilitated by FLUKA.

The BeAGLE model categorizes the $e$+$A$ collision process into three distinct stages: hard scattering processes, IntraNuclear Cascade (INC) processes~\cite{Cugnon:1982qw}, and the breakup of excited nuclei~\cite{Serber:1947zz, Monira:2023}. Within the model framework, these stages occur sequentially, each on distinct time scales in the nuclear rest frame. Hard scattering processes happen almost instantaneously during the collision, while INC processes and the subsequent breakup of excited nuclei occur over time scales of approximately \(10^{-22}\) and \(10^{-16}\) seconds, respectively~\cite{Serber:1947zz, Monira:2023}. Therefore, isolating the kinematic regions where each process predominates is essential for our study. Thus, it's important to look first into the $dN/d\eta$ in $e$+$A$ collisions given in Fig.~\ref{fig:1} panel (b).

The $\eta$ acceptance used in this work is inspired by the ePIC experiment at the EI~\cite{AbdulKhalek:2021gbh, ePIC} as (i) $-3.5 < \eta < 3.5$, (ii) $4.5< \eta < 5.9$ and (iii) $\eta > 6.0$. Figure~\ref{fig:1} presents the $dN/d\eta$ distribution of the $e$+$Au$ collisions at 18$\times$110 (GeV) from the BeAGLE model with the three $\eta$-regions acceptance. 
In this work, we investigate the potential of using (i) the mean energy ($\langle E \rangle$), (ii) mean transverse momentum ($\langle p_T \rangle$), and (iii) the mean multiplicity ($\langle N \rangle$). The $\langle E \rangle$ and $\langle p_T \rangle$ are given as;
\begin{eqnarray}
     \langle E \rangle   &=& \frac{\sum_{i} w_{i} E_{i}}{\sum_{i} w_{i}}, ~ \langle p_{T} \rangle = \frac{\sum_{i} w_{i} p_{T_{i}}}{\sum_{i} w_{i}},
\end{eqnarray}
where the sum runs over the particle in the used acceptance, the $w_{i}$ is the i$^{th}$ particle weight that considers the detector acceptance and efficiency effects. For a perfect detector or MC model, $w_{i} = 1.0$.
 
\subsection{Results and discussion}\label{sec:3}
In the context of ion-ion collisions~\cite{Busza:2018rrf}, geometric properties such as the impact parameter \(b\) strongly correlate with the number of particles produced in the collision. Consequently, the experimental measurements of number multiplicity can be used for centrality estimation~\cite{Busza:2018rrf}. In this work, we are looking into the $b$ correlations with (i) $\langle E \rangle$, (ii) $\langle p_T \rangle$, and (iii) $\langle N \rangle$ for the three $\eta$ regions highlighted in Fig~\ref{fig:1} (b). Such correlations can reflect the ability of the produced particles in e+A collisions to be used for a centrality definition at the EIC. 

 \begin{figure*}[!h]
 \centering{
 \includegraphics[width=0.99\linewidth,angle=0]{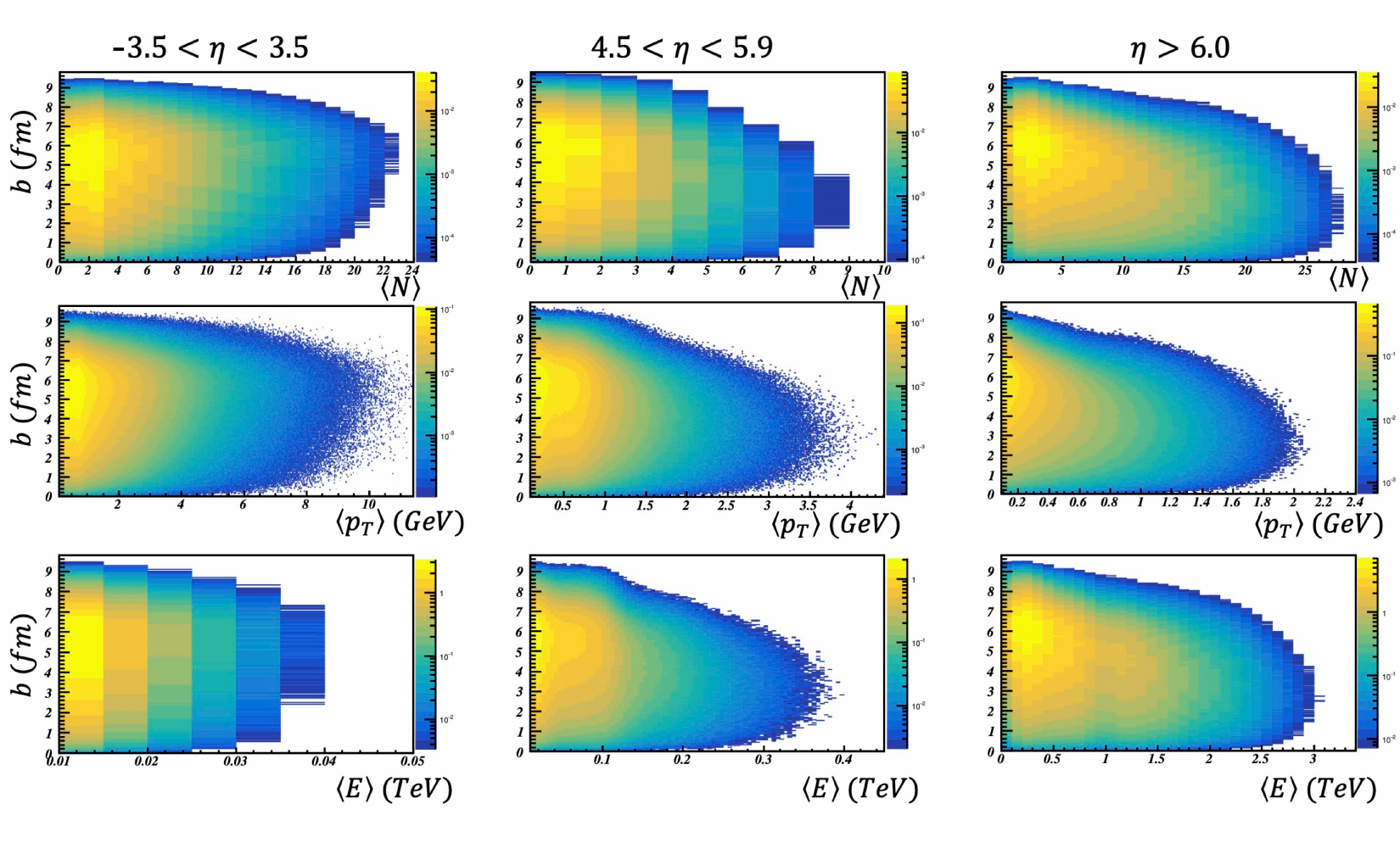}
\vskip -0.36cm
 \caption{
 The impact parameter correlations with $\langle N \rangle$, $\langle p_T \rangle$, and $\langle E \rangle$ from the  $e$+$Au$ collisions at 18$\times$110 (GeV) from the BeAGLE model. The correlations are presented for the three $\eta$ regions given in Fig~\ref{fig:1}
 \label{fig:2}
 }
 }
 \end{figure*}
Figure~\ref{fig:2} presents the impact parameter correlations with the $\langle N \rangle$, $\langle p_T \rangle$, and $\langle E \rangle$, for $e$+$Au$ collisions at 18$\times$110 (GeV) from the BeAGLE model with the three indicated $\eta$-regions acceptance.  Our results indicated similarities in correlations between the three proposed observables for the same $\eta$ acceptance. In addition, the $-3.5 < \eta < 3.5$ results show a weak correlation, if any, between b and $\langle N \rangle$, $\langle E \rangle$, and  $\langle p_T \rangle$.  In contrast, results with $4.5$ $<$ $\eta$ $<$ $5.9$ and $\eta$ $>$ $6.0$ show modest correlations, in agreement with prior investigation~\cite{Chang:2022hkt}. Consequently, quantifying such correlations' ability to classify the centrality, we will project the b-distribution for two cuts over large and small $\langle E \rangle$ values.

 \begin{figure*}[!h]
 \centering{
 \includegraphics[width=0.99\linewidth,angle=0]{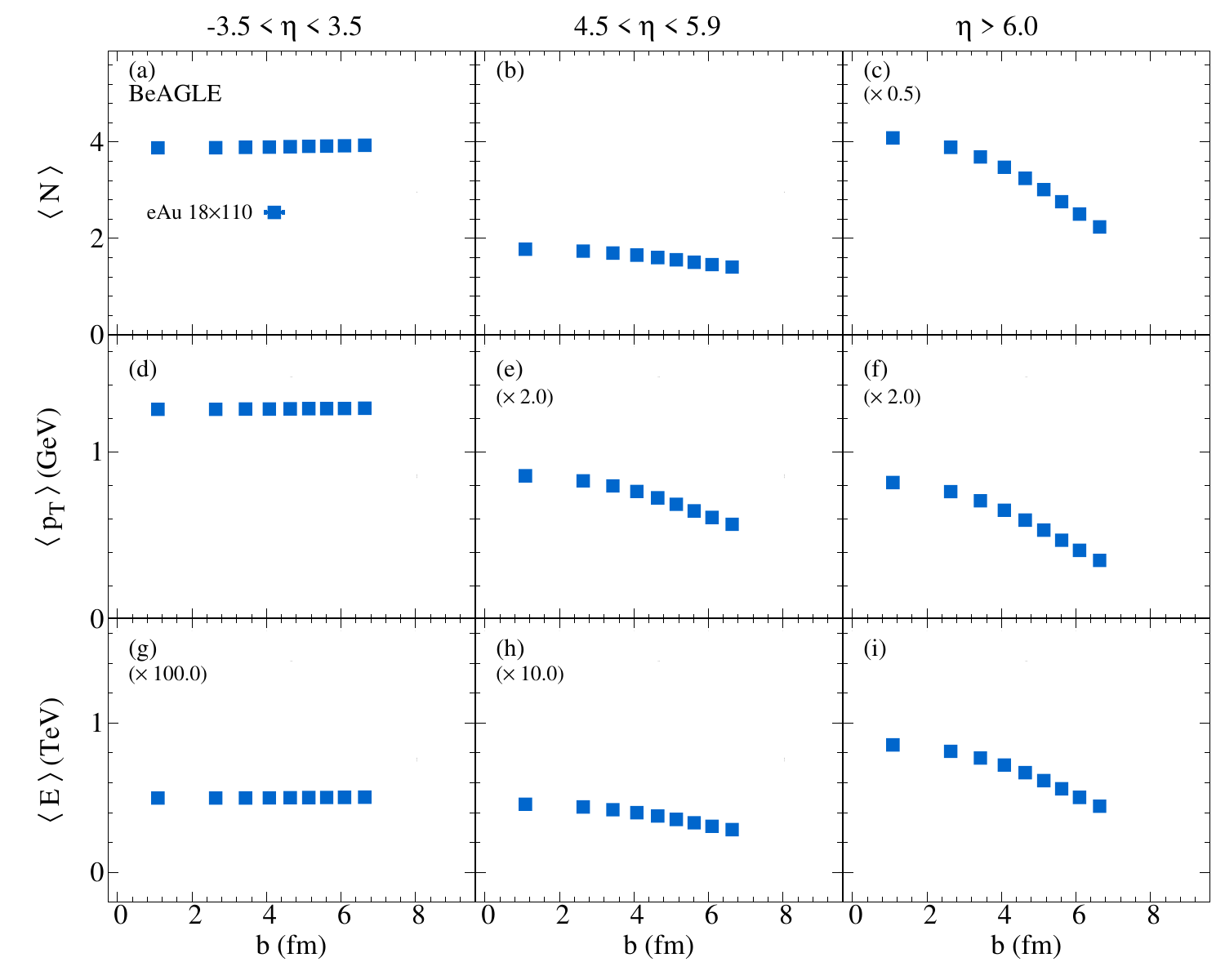}
\vskip -0.36cm
 \caption{
 The impact parameter dependence of the $\langle N \rangle$, $\langle p_T \rangle$, and $\langle E \rangle$ for $e$+$Au$ collisions at 18$\times$110 (GeV) from the BeAGLE model. The correlations are presented for the three $\eta$ regions in Fig~\ref{fig:1}.
 \label{fig:3}
 }
 }
 \end{figure*}
Reducing the two-dimensional correlations into one-dimensional mean values for different impact parameter ($b$) selections can provide deeper insights into the data presented in Fig.~\ref{fig:2}. Figure~\ref{fig:3} illustrates the impact parameter dependence of the mean number of particles ($\langle N \rangle$), mean transverse momentum ($\langle p_T \rangle$), and mean energy ($\langle E \rangle$) for $e$+$Au$ collisions at 18$\times$110 GeV, based on the BeAGLE model framework (see Appendix~\ref{app:a} for e+Au at 10$\times$40, and e+Ru at 18$\times$110). Our mid-rapidity results, shown in panels (a), (d), and (g), indicate that $\langle N \rangle$, $\langle p_T \rangle$, and $\langle E \rangle$ are largely insensitive to variations in the impact parameter.
However, forward and far-forward rapidity results exhibit an apparent sensitivity to impact parameter changes, with mean values varying by approximately 30\% and 50\%, respectively.
This observation suggests the need for further investigation into the nature of impact parameter distributions across different centrality selections.

 \begin{figure}[!h]
 \centering{
 \includegraphics[width=0.99\linewidth,angle=0]{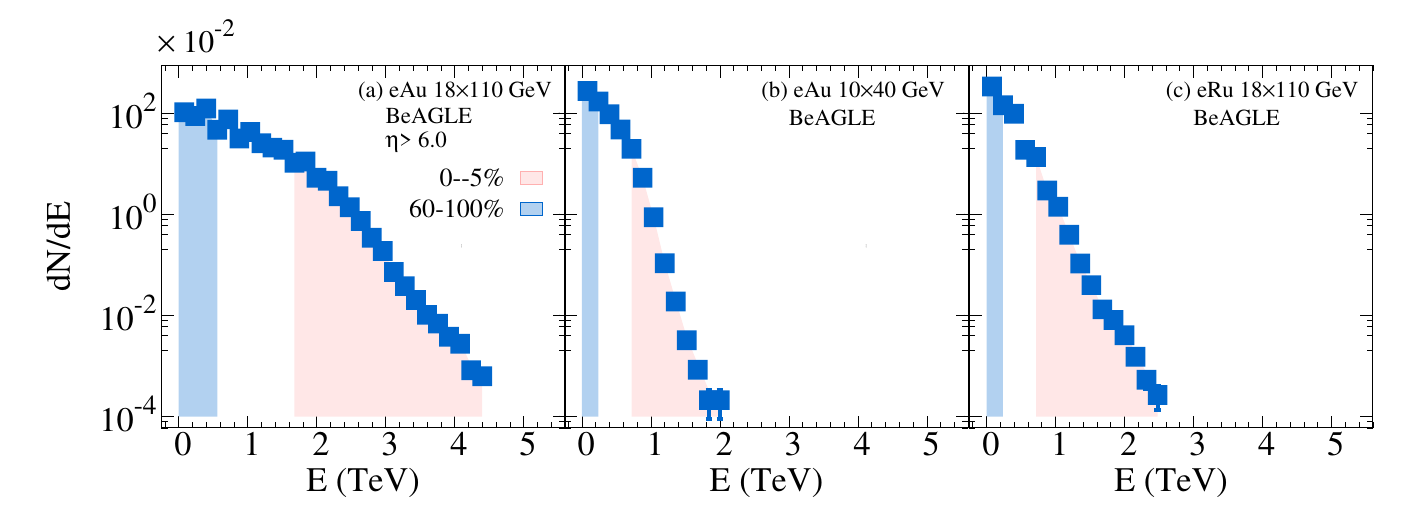}
\vskip -0.36cm
 \caption{
 The energy distributions at $\eta > 6.0$ for (a) e+Au at 18$\times$110,  (b) e+Au at 10$\times$40, and (c) e+Ru at 18$\times$110 from the BeAGLE model. The shaded band represents the 0--5\% and 60-100\% selection on the energy distributions.
 \label{fig:4}
 }
 }
 \end{figure}
To quantify the ability of the $\langle E \rangle$ to constrain the centrality in e+A collisions, we will need first to present the $\langle E \rangle$ distribution and then impose percentile cuts on it. Figure~\ref{fig:4} presents an example of the $\langle E \rangle$ distribution for $\eta > 6.0$. The results are presented for  e+Au at 18$\times$110, e+Au at 10$\times$40, and e+Ru at 18$\times$110 collisions in panels (a), (b) and (c) respectively. Our results show that reducing the collision energy or nuclear size reduces the energy deposited at far forward rapidity by about 50\%. The presented bands in Fig.~\ref{fig:4} represent the 0--5\% and 60-100\% selections on the energy distributions.

 \begin{figure*}[!h]
 \centering{
 \includegraphics[width=0.99\linewidth,angle=0]{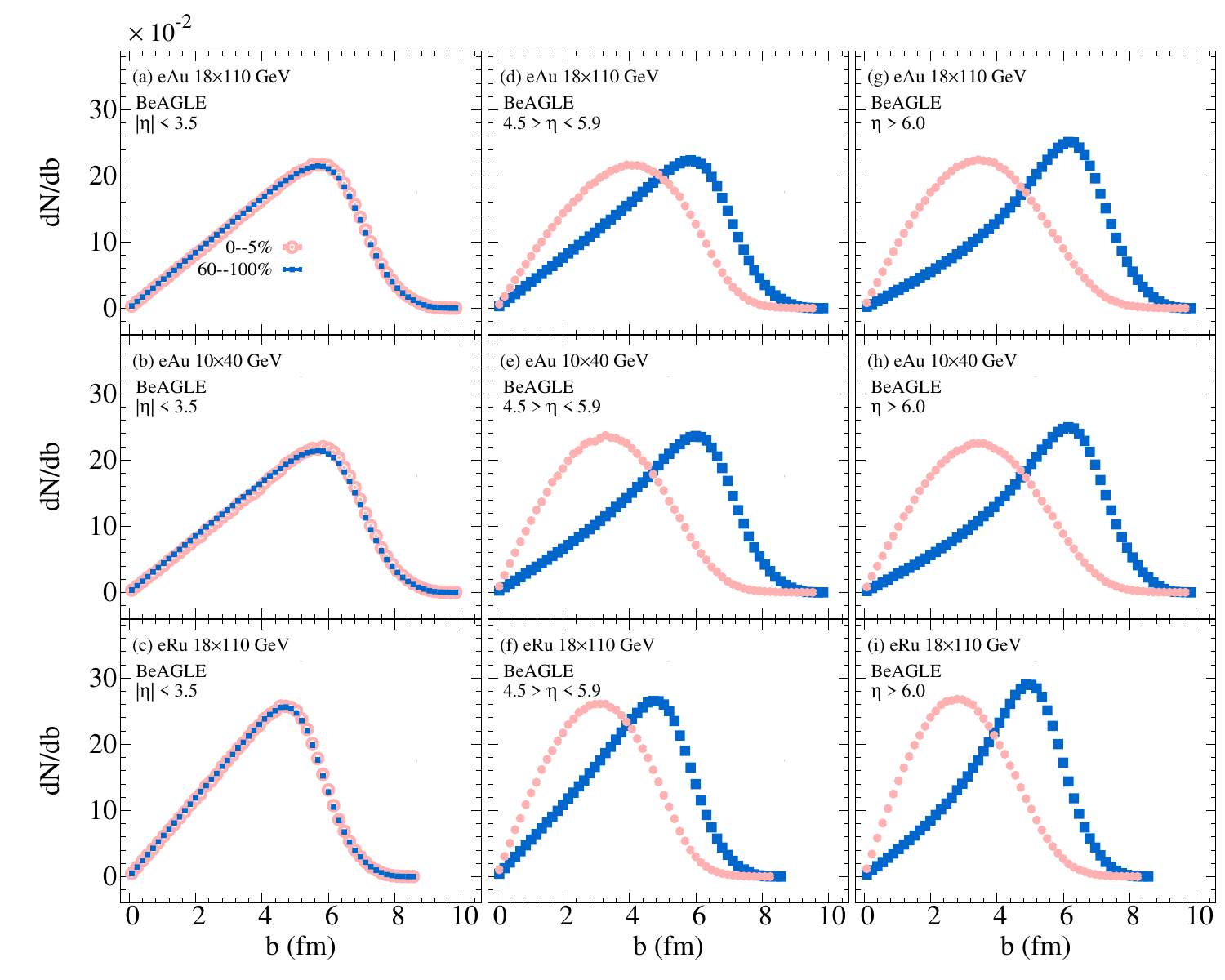}
\vskip -0.36cm
 \caption{
 The impact parameter distributions with $|\eta|$ $<$ $3.5$, $4.5$ $<$ $\eta$ $<$ $5.9$ and $\eta$ $>$ $6.0$ for e+Au at 18$\times$110,  e+Au at 10$\times$40, and e+Ru at 18$\times$110 collisions from the BeAGLE model.
 \label{fig:5}
 }
 }
 \end{figure*}
The impact parameter distribution for the 0--5\% and 60-100\% selections, more selections are presented in Appendix~\ref{app:a}, from the BeAGLE model at different collision energies and system sizes are presented in Fig.~\ref{fig:5}. The results are presented for the three $\eta$ selections indicated in Fig.~\ref{fig:1} (b). For mid-rapidity acceptance, no difference is observed between central and peripheral collisions for all energies and systems presented. In contrast, forward and far forward rapidity selection results indicated visible differences between central and peripheral collisions. We observed about 30\% and 50\% deference in the mean value of the distributions between central and peripheral collisions for the forward and far forward rapidity selection, respectively. In addition, our results indicated that the differences have a fragile sensitivity to collision energies and system size variations. 
Consequently, it's essential to point out that while we observe apparent differences in the impact parameter distribution mean values, the distributions are broad and have large intercept areas between central and peripheral collisions. Such observation suggests that using the forward and far forward energy as an experimental handle on the collision geometry is complicated.

Our results in Figs.~\ref{fig:3} and~\ref{fig:5} show that event properties such as $\langle N \rangle$, $\langle p_T \rangle$, and $\langle E \rangle$ at mid-rapidity exhibit no dependence on the impact parameter. This can be explained by Fig.~\ref{fig:1} (b), which suggests that mid-rapidity particles are predominantly produced by hard scattering processes. In the BeAGLE model, these hard scattering processes are simulated using the PYTHIA-6 model, which does not account for the impact parameter.
In contrast, the $\langle N \rangle$, $\langle p_T \rangle$, and $\langle E \rangle$ simulated at forward and far-forward rapidity show modest differences between central and peripheral collisions. As seen in Fig.~\ref{fig:1} (b), these differences reflect the impact parameter dependence of intranuclear cascade processes at forward rapidity and the breakup of excited nuclear remnants at far-forward rapidity. Our findings suggest that further investigations using alternative models for (i) intranuclear cascade and (ii) nuclear remnant breakup could provide deeper insights into the relationship between these processes and the impact parameter.

\section{Conclusion}\label{sec:4}
In this study, we investigated the feasibility of defining centrality in e+A collisions at the EIC by examining correlations between the impact parameter and several observables: total energy, total transverse momentum, and total number of particles. Our analysis focused on three distinct kinematic regions: mid-rapidity, forward rapidity, and far-forward rapidity. The results reveal that the correlation between the impact parameter and these observables is weak in the mid-rapidity region, indicating no significant difference between centrality classes (0-5\% and 60-100\%). However, we observed a more pronounced difference between central and peripheral collisions in the forward and far-forward rapidity regions, suggesting a modest correlation between the impact parameter and the observables in these regions. Despite this observed difference, the correlation is not sufficiently robust to definitively define centrality in electron-ion collisions. The broad impact parameter distributions and substantial overlap between central and peripheral collisions in these regions pose challenges for precise centrality determination.

Our findings underscore the inherent complexity of defining centrality in electron-ion collisions compared to ion-ion collisions. While centrality is well-established in the context of ion-ion collisions, defining it for electron-ion collisions remains an open question. Further investigation is necessary to develop a robust definition of centrality for the EIC. This may involve exploring additional variables or refining the existing correlations to achieve a more precise characterization of centrality in electron-ion collisions. The advent of actual experiments at the EIC will be instrumental in validating or refining our current understanding and potentially uncovering new avenues for defining centrality in this context.

\section*{Acknowledgments}
The authors thank Roy Lacey and Tanner Mengel for the valuable discussions and for pointing out essential references. This work was supported in part by funding from the Division of Nuclear Physics of the U.S. Department of Energy under Grant No. DE-FG02-96ER40982 (NM). AR and AH acknowledge the support of the American University in Cairo under the agreement Number SSE-PHYS-A.H-FY24-RG-2023-Dec-17-14-59-41.


\appendix

\section{}
\label{app:a}
In this section, we present additional figures that provide further details on the dependence of $\langle N \rangle$, $\langle p_T \rangle$, and $\langle E \rangle$ on the impact parameter.
 \begin{figure*}[!h]
 \centering{
 \includegraphics[width=0.62\linewidth,angle=0]{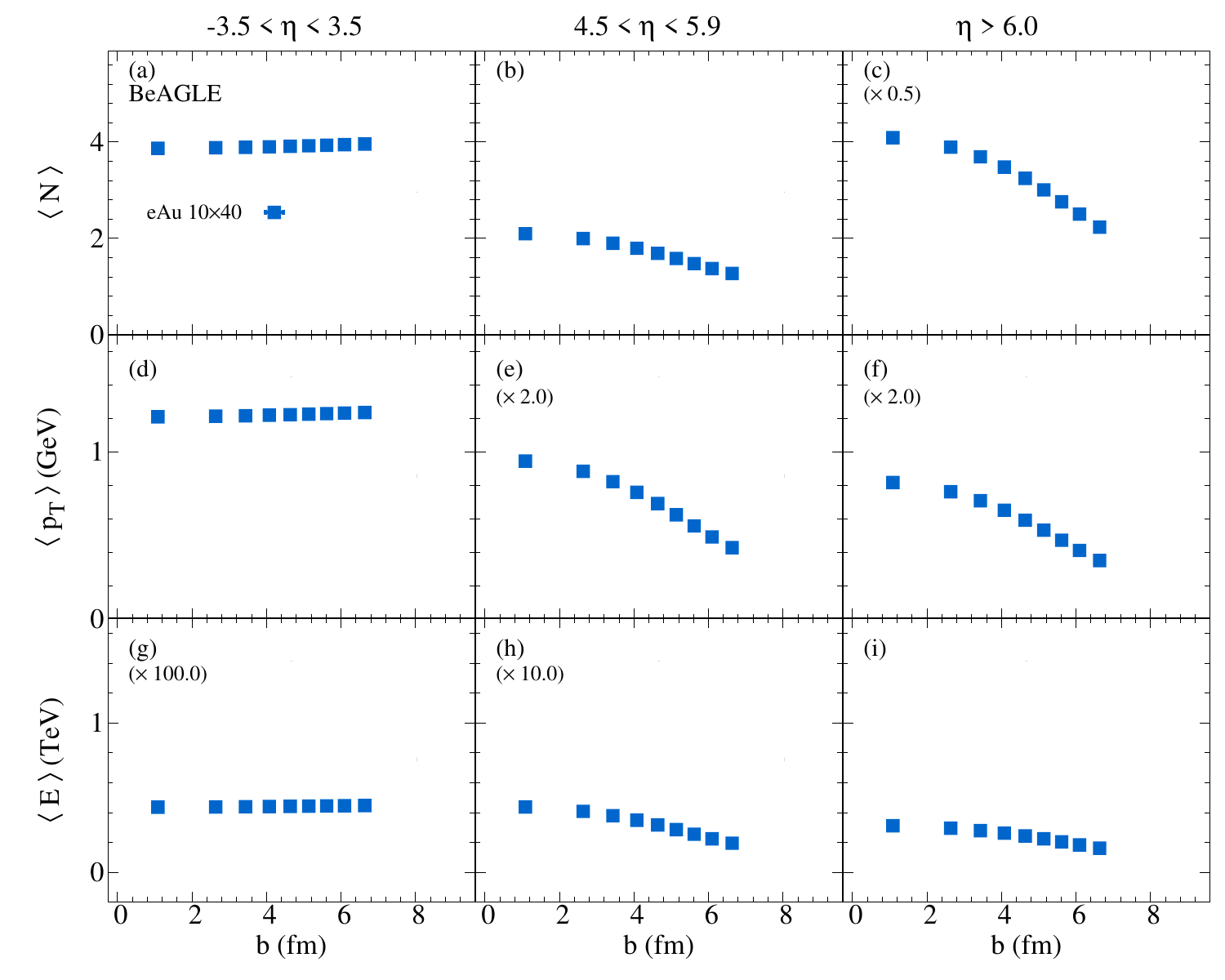}
\vskip -0.36cm
 \caption{
 The impact parameter dependence of the $\langle N \rangle$, $\langle p_T \rangle$, and $\langle E \rangle$ for $e$+$Au$ collisions at 10$\times$40 (GeV) from the BeAGLE model. The correlations are presented for the three $\eta$ regions in Fig~\ref{fig:1}.
 \label{fig:4}
 }
 }
 \end{figure*}
 \begin{figure*}[!h]
 \centering{
 \includegraphics[width=0.62\linewidth,angle=0]{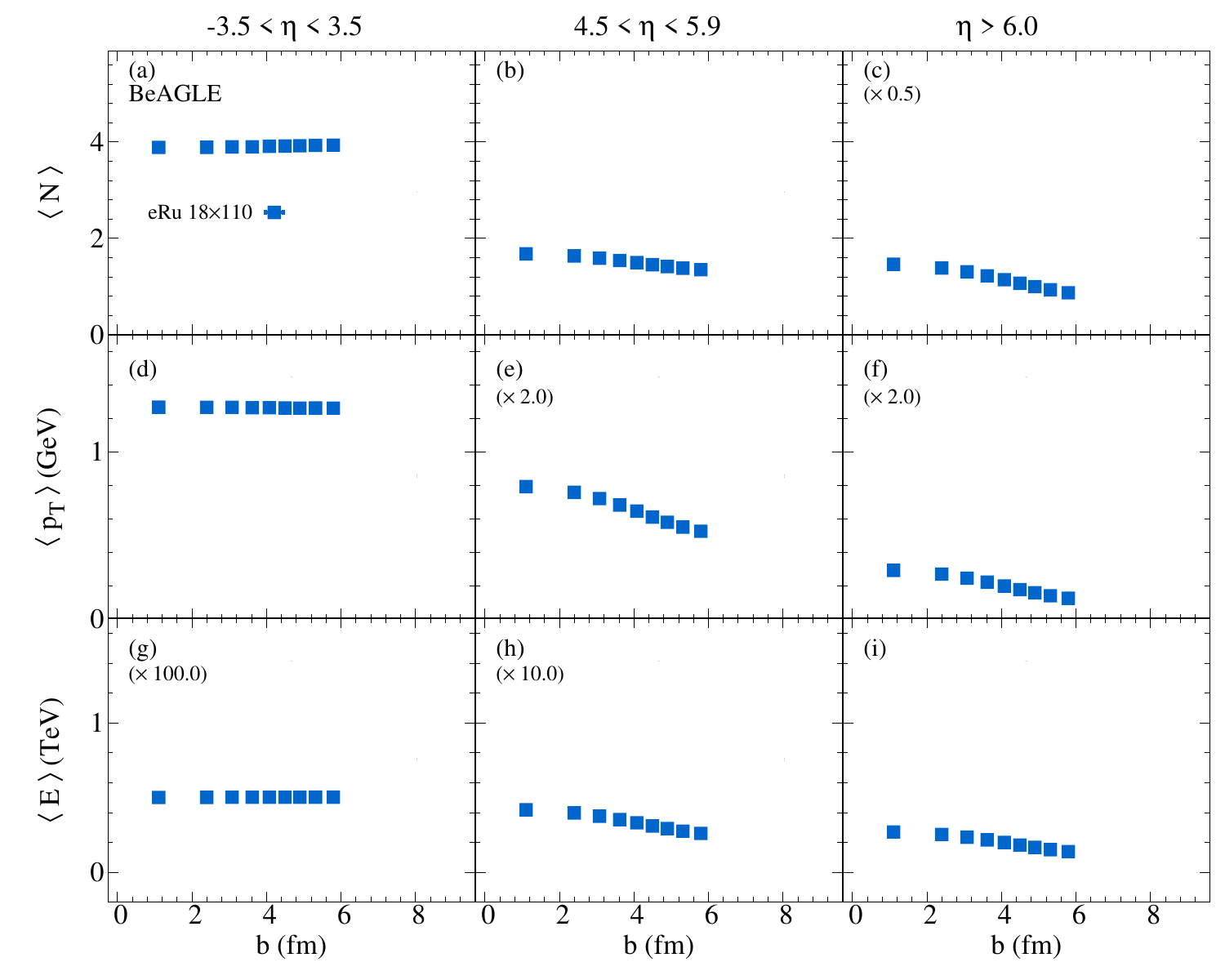}
\vskip -0.36cm
 \caption{
 The impact parameter dependence of the $\langle N \rangle$, $\langle p_T \rangle$, and $\langle E \rangle$ for $e$+$Ru$ collisions at 18$\times$110 (GeV) from the BeAGLE model. The correlations are presented for the three $\eta$ regions in Fig~\ref{fig:1}.
 \label{fig:4}
 }
 }
 \end{figure*}
 \begin{figure*}[!h]
 \centering{
 \includegraphics[width=0.65\linewidth,angle=0]{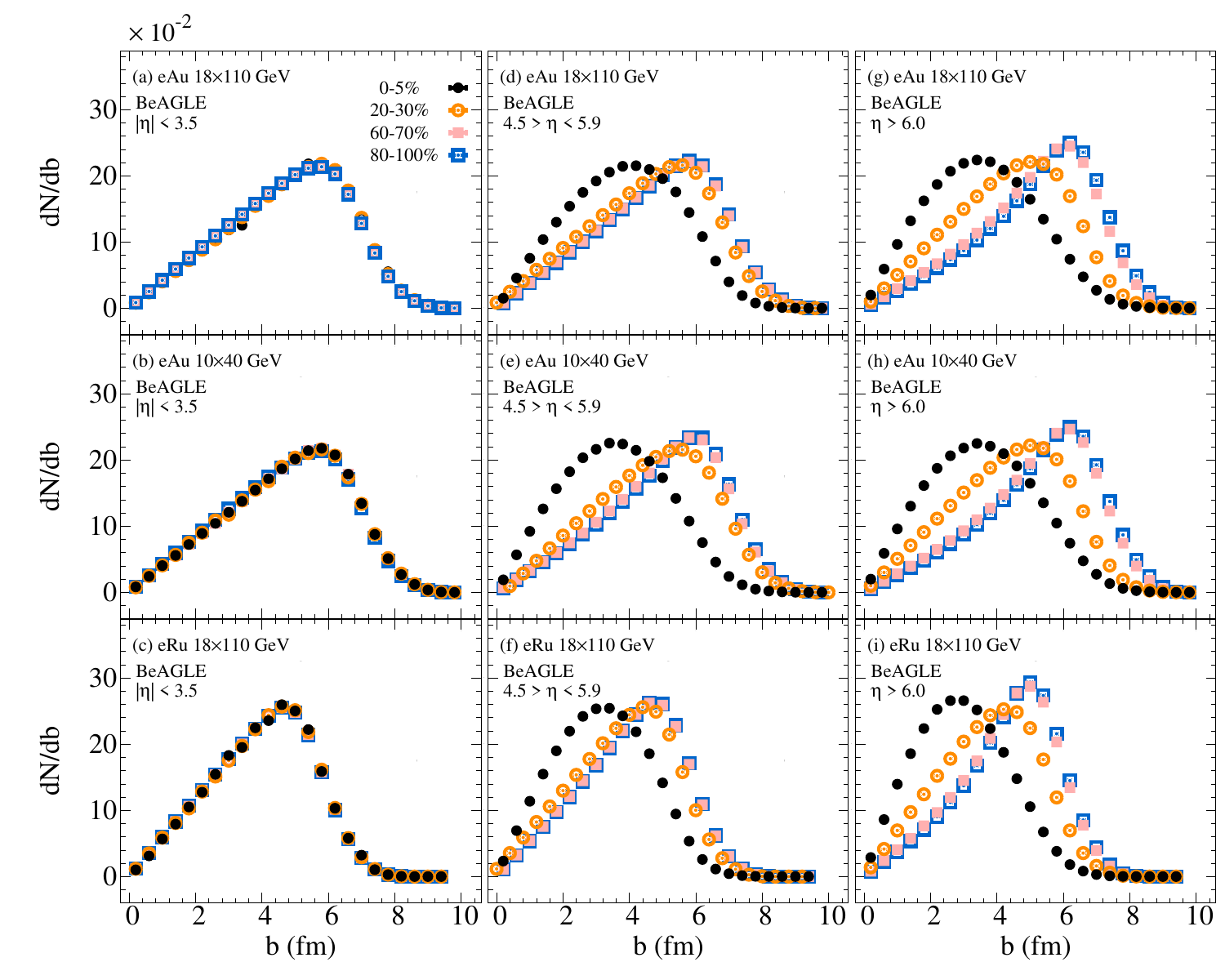}
\vskip -0.36cm
 \caption{
 The impact parameter correlations with $\langle N \rangle$, $\langle p_T \rangle$, and $\langle E \rangle$ from the  $e$+$Au$ collisions at 18$\times$110 (GeV) from the BeAGLE model. The correlations are presented for the three $\eta$ regions given in Fig~\ref{fig:1}
 \label{fig:4}
 }
 }
 \end{figure*}

\bibliography{ref}

\begin{thebibliography}{16}
\expandafter\ifx\csname natexlab\endcsname\relax\def\natexlab#1{#1}\fi
\expandafter\ifx\csname bibnamefont\endcsname\relax
  \def\bibnamefont#1{#1}\fi
\expandafter\ifx\csname bibfnamefont\endcsname\relax
  \def\bibfnamefont#1{#1}\fi
\expandafter\ifx\csname citenamefont\endcsname\relax
  \def\citenamefont#1{#1}\fi
\expandafter\ifx\csname url\endcsname\relax
  \def\url#1{\texttt{#1}}\fi
\expandafter\ifx\csname urlprefix\endcsname\relax\def\urlprefix{URL }\fi
\providecommand{\bibinfo}[2]{#2}
\providecommand{\eprint}[2][]{\url{#2}}

\bibitem[{\citenamefont{Krane}(1988)}]{Krane:359790}
\bibinfo{author}{\bibfnamefont{K.~S.} \bibnamefont{Krane}},
  \emph{\bibinfo{title}{{Introductory nuclear physics}}}
  (\bibinfo{publisher}{Wiley}, \bibinfo{address}{New York, NY},
  \bibinfo{year}{1988}), \urlprefix\url{https://cds.cern.ch/record/359790}.

\bibitem[{\citenamefont{Miller et~al.}(2007)\citenamefont{Miller, Reygers,
  Sanders, and Steinberg}}]{Miller:2007ri}
\bibinfo{author}{\bibfnamefont{M.~L.} \bibnamefont{Miller}},
  \bibinfo{author}{\bibfnamefont{K.}~\bibnamefont{Reygers}},
  \bibinfo{author}{\bibfnamefont{S.~J.} \bibnamefont{Sanders}},
  \bibnamefont{and}
  \bibinfo{author}{\bibfnamefont{P.}~\bibnamefont{Steinberg}},
  \bibinfo{journal}{Ann. Rev. Nucl. Part. Sci.} \textbf{\bibinfo{volume}{57}},
  \bibinfo{pages}{205} (\bibinfo{year}{2007}), \eprint{nucl-ex/0701025}.

\bibitem[{\citenamefont{Abelev et~al.}(2013)}]{ALICE:2013hur}
\bibinfo{author}{\bibfnamefont{B.}~\bibnamefont{Abelev}} \bibnamefont{et~al.}
  (\bibinfo{collaboration}{ALICE}), \bibinfo{journal}{Phys. Rev. C}
  \textbf{\bibinfo{volume}{88}}, \bibinfo{pages}{044909}
  (\bibinfo{year}{2013}), \eprint{1301.4361}.

\bibitem[{\citenamefont{Abdul~Khalek et~al.}(2022)}]{AbdulKhalek:2021gbh}
\bibinfo{author}{\bibfnamefont{R.}~\bibnamefont{Abdul~Khalek}}
  \bibnamefont{et~al.}, \bibinfo{journal}{Nucl. Phys. A}
  \textbf{\bibinfo{volume}{1026}}, \bibinfo{pages}{122447}
  (\bibinfo{year}{2022}), \eprint{2103.05419}.

\bibitem[{\citenamefont{Chang et~al.}(2022)\citenamefont{Chang, Aschenauer,
  Baker, Jentsch, Lee, Tu, Yin, and Zheng}}]{Chang:2022hkt}
\bibinfo{author}{\bibfnamefont{W.}~\bibnamefont{Chang}},
  \bibinfo{author}{\bibfnamefont{E.-C.} \bibnamefont{Aschenauer}},
  \bibinfo{author}{\bibfnamefont{M.~D.} \bibnamefont{Baker}},
  \bibinfo{author}{\bibfnamefont{A.}~\bibnamefont{Jentsch}},
  \bibinfo{author}{\bibfnamefont{J.-H.} \bibnamefont{Lee}},
  \bibinfo{author}{\bibfnamefont{Z.}~\bibnamefont{Tu}},
  \bibinfo{author}{\bibfnamefont{Z.}~\bibnamefont{Yin}}, \bibnamefont{and}
  \bibinfo{author}{\bibfnamefont{L.}~\bibnamefont{Zheng}},
  \bibinfo{journal}{Phys. Rev. D} \textbf{\bibinfo{volume}{106}},
  \bibinfo{pages}{012007} (\bibinfo{year}{2022}), \eprint{2204.11998}.

\bibitem[{\citenamefont{Magdy et~al.}(2024)\citenamefont{Magdy, Hegazy, Rafaat,
  Li, Deshpande, Abdelhady, Ellithi, Lacey, and Tu}}]{Magdy:2024thf}
\bibinfo{author}{\bibfnamefont{N.}~\bibnamefont{Magdy}},
  \bibinfo{author}{\bibfnamefont{M.}~\bibnamefont{Hegazy}},
  \bibinfo{author}{\bibfnamefont{A.}~\bibnamefont{Rafaat}},
  \bibinfo{author}{\bibfnamefont{W.}~\bibnamefont{Li}},
  \bibinfo{author}{\bibfnamefont{A.}~\bibnamefont{Deshpande}},
  \bibinfo{author}{\bibfnamefont{A.~M.~H.} \bibnamefont{Abdelhady}},
  \bibinfo{author}{\bibfnamefont{A.~Y.} \bibnamefont{Ellithi}},
  \bibinfo{author}{\bibfnamefont{R.~A.} \bibnamefont{Lacey}}, \bibnamefont{and}
  \bibinfo{author}{\bibfnamefont{Z.}~\bibnamefont{Tu}} (\bibinfo{year}{2024}),
  \eprint{2405.07844}.

\bibitem[{\citenamefont{Li et~al.}(2024)\citenamefont{Li, Liu, and
  Vitev}}]{Li:2023dhb}
\bibinfo{author}{\bibfnamefont{H.~T.} \bibnamefont{Li}},
  \bibinfo{author}{\bibfnamefont{Z.~L.} \bibnamefont{Liu}}, \bibnamefont{and}
  \bibinfo{author}{\bibfnamefont{I.}~\bibnamefont{Vitev}},
  \bibinfo{journal}{Phys. Lett. B} \textbf{\bibinfo{volume}{848}},
  \bibinfo{pages}{138354} (\bibinfo{year}{2024}), \eprint{2303.14201}.

\bibitem[{\citenamefont{Piller et~al.}(1995)\citenamefont{Piller, Ratzka, and
  Weise}}]{Piller:1995kh}
\bibinfo{author}{\bibfnamefont{G.}~\bibnamefont{Piller}},
  \bibinfo{author}{\bibfnamefont{W.}~\bibnamefont{Ratzka}}, \bibnamefont{and}
  \bibinfo{author}{\bibfnamefont{W.}~\bibnamefont{Weise}}, \bibinfo{journal}{Z.
  Phys. A} \textbf{\bibinfo{volume}{352}}, \bibinfo{pages}{427}
  (\bibinfo{year}{1995}), \eprint{hep-ph/9504407}.

\bibitem[{\citenamefont{Cugnon}(1982)}]{Cugnon:1982qw}
\bibinfo{author}{\bibfnamefont{J.}~\bibnamefont{Cugnon}},
  \bibinfo{journal}{Nucl. Phys. A} \textbf{\bibinfo{volume}{387}},
  \bibinfo{pages}{191C} (\bibinfo{year}{1982}).

\bibitem[{\citenamefont{Weisskopf}(1937)}]{Weisskopf:1937zz}
\bibinfo{author}{\bibfnamefont{V.}~\bibnamefont{Weisskopf}},
  \bibinfo{journal}{Phys. Rev.} \textbf{\bibinfo{volume}{52}},
  \bibinfo{pages}{295} (\bibinfo{year}{1937}).

\bibitem[{\citenamefont{Serber}(1947)}]{Serber:1947zz}
\bibinfo{author}{\bibfnamefont{R.}~\bibnamefont{Serber}},
  \bibinfo{journal}{Phys. Rev.} \textbf{\bibinfo{volume}{72}},
  \bibinfo{pages}{1008} (\bibinfo{year}{1947}).

\bibitem[{\citenamefont{Monira}(2023)}]{Monira:2023}
\bibinfo{author}{\bibfnamefont{J.}~\bibnamefont{Monira}}, Ph.D. thesis,
  \bibinfo{school}{Graduate School of Engineering, Kyushu University, Japan}
  (\bibinfo{year}{2023}).

\bibitem[{\citenamefont{Mathews et~al.}(1982)\citenamefont{Mathews, Glagola,
  Moyle, and Viola}}]{Mathews:1982zz}
\bibinfo{author}{\bibfnamefont{G.~J.} \bibnamefont{Mathews}},
  \bibinfo{author}{\bibfnamefont{B.~G.} \bibnamefont{Glagola}},
  \bibinfo{author}{\bibfnamefont{R.~A.} \bibnamefont{Moyle}}, \bibnamefont{and}
  \bibinfo{author}{\bibfnamefont{V.~E.} \bibnamefont{Viola}},
  \bibinfo{journal}{Phys. Rev. C} \textbf{\bibinfo{volume}{25}},
  \bibinfo{pages}{2181} (\bibinfo{year}{1982}).

\bibitem[{\citenamefont{Salgado and Wiedemann}(2003)}]{SW:2003}
\bibinfo{author}{\bibfnamefont{C.}~\bibnamefont{Salgado}} \bibnamefont{and}
  \bibinfo{author}{\bibfnamefont{U.~A.} \bibnamefont{Wiedemann}},
  \bibinfo{journal}{Phys. Rev. D} \textbf{\bibinfo{volume}{68}},
  \bibinfo{pages}{014008} (\bibinfo{year}{2003}), \eprint{hep-ph/0302184}.

\bibitem[{ePI()}]{ePIC}
\emph{\bibinfo{title}{The epic collaboration}},
  \bibinfo{howpublished}{\url{https://www.epic-eic.org/}}.

\bibitem[{\citenamefont{Busza et~al.}(2018)\citenamefont{Busza, Rajagopal, and
  van~der Schee}}]{Busza:2018rrf}
\bibinfo{author}{\bibfnamefont{W.}~\bibnamefont{Busza}},
  \bibinfo{author}{\bibfnamefont{K.}~\bibnamefont{Rajagopal}},
  \bibnamefont{and} \bibinfo{author}{\bibfnamefont{W.}~\bibnamefont{van~der
  Schee}}, \bibinfo{journal}{Ann. Rev. Nucl. Part. Sci.}
  \textbf{\bibinfo{volume}{68}}, \bibinfo{pages}{339} (\bibinfo{year}{2018}),
  \eprint{1802.04801}.

\end{thebibliography}

\end{document}